\documentclass[aps,prl,showkeys,tightenlines,nofootinbib, twocolumn]{revtex4-1}
\usepackage{comment}

\usepackage{amsmath,amssymb, amsfonts}
\usepackage{graphicx}
\usepackage[usenames, dvipsnames]{color}

\newcommand{\bea}{\begin{eqnarray}}
\newcommand{\eea}{\end{eqnarray}}
\newcommand{\nn}{\nonumber}

\begin{document}

\title{NRQCD Confronts LHCb Data on Quarkonium Production within Jets} 
 
\author{Reggie Bain}
\email{rab59@duke.edu}
\affiliation{Department of Physics, Duke University, Durham, NC, 27713, USA}

\author{Yiannis Makris}
\email{yiannis.makris@duke.edu}
\affiliation{Department of Physics, Duke University, Durham, NC, 27713, USA}

\author{Thomas Mehen}
\email{mehen@phy.duke.edu}
\affiliation{Department of Physics, Duke University, Durham, NC, 27713, USA}

\author{Lin Dai}
\email{lid33@pitt.edu}
\affiliation{Pittsburgh Particle Physics Astrophysics and Cosmology Center (PITT PACC), Department of Physics and Astronomy, University of Pittsburgh, 3941 O'Hara St., Pittsburgh, PA 15260, USA}

\author{Adam K. Leibovich}
\email{akl2@pitt.edu}
\affiliation{Pittsburgh Particle Physics Astrophysics and Cosmology Center (PITT PACC), Department of Physics and Astronomy, University of Pittsburgh, 3941 O'Hara St., Pittsburgh, PA 15260, USA}


\begin{abstract}
\noindent
We analyze the recent LHCb measurement of the distribution of the fraction of the transverse momentum, $z(J/\psi)$, 
carried by the $J/\psi$ within a jet. LHCb data is compared to analytic calculations using the fragmenting jet function
(FJF) formalism for studying $J/\psi$ in jets.  Logarithms in the FJFs are resummed using DGLAP evolution. We also 
convolve hard QCD partonic cross sections, showered with
PYTHIA, with leading order Non-Relativistic Quantum Chromodynamics (NRQCD)
fragmentation functions and obtain consistent results.  Both approaches use Madgraph to calculate the hard process that 
creates the jet initiating parton. These calculations give reasonable agreement with  the $z(J/\psi)$ distribution that
was shown to be poorly described by default PYTHIA simulations in the LHCb paper.  
We compare our predictions for the $J/\psi$ distribution using various extractions of nonperturbative NRQCD 
long-distance matrix elements (LDMEs) in the literature. NRQCD calculations agree with LHCb data better than 
default PYTHIA regardless of which fit to the LDMEs is used.
 LDMEs from fits that focus exclusively on high transverse 
momentum data from colliders are in good agreement with the LHCb measurement. 

\end{abstract}



\keywords{}


\maketitle


The production of quarkonium is
a  challenging test  of Quantum Chromodynamics due to the mutiple length scales involved. The LHCb collaboration \cite{Aaij:2017fak}  published the first study 
of $J/\psi$ produced within jets. The distribution of the fraction of the jet's transverse momentum,  $p_T$, carried 
by the $J/\psi$, $z(J/\psi)$, was found to disagree significantly with predictions from the PYTHIA monte carlo~\cite{Sjostrand:2006za,Sjostrand:2014zea} using  leading order calculations 
of $J/\psi$ production in the Non-Relativistic Quantum Chromodynamics (NRQCD) factorization formalism~\cite{Bodwin:1994jh}. This letter is provides improved theoretical 
calculations of the $z(J/\psi)$ distribution and to discuss the implications of the LHCb results for the NRQCD factorization formalism. 

Production of quarkonium in hadron colliders has been the subject of experimental and theoretical studies for decades.
The problem is challenging because it involves several disparate scales. These include $p_T$, which can be much larger than the mass 
of the bound state, $â\approx 2m_Q$, where $m_Q$ is the mass of the heavy quark, as well as scales that are much smaller: the relative momenta, 
$m_Q v$ ($v$ is the typical velocity of the heavy quarks in the bound state), the kinetic energy, $m_Qv^2$, and the nonperturbative scale $\Lambda_{QCD}$.
 
The most common approach to 
calculating quarkonium production is the NRQCD factorization formalism~\cite{Bodwin:1994jh}. In this formalism, the cross section 
for $J/\psi$ in a $pp$ collision is written as 
\bea
d\sigma[pp\to J/\psi X] =  \sum_n d\sigma[pp\to c \bar{c}(n) X] \langle {\cal O}^{J/\psi}(n)\rangle, \nn
\eea
where $d\sigma[pp\to c \bar{c}(n)X]$ is the short distance cross section for producing the $c\bar{c}$ pair in a state $n$ with definite color and angular momentum quantum numbers and $\langle {\cal O}^{J/\psi}(n)\rangle$ is a long distance matrix element (LDME) that describes the nonperturbative transition of the $c\bar{c}$ pair in the state $n$ into a final state containing $J/\psi$.  $X$ denotes other possible particles in the final state. The quantum numbers $n$ will be denoted $^{2S+1}L_J^{[i]}$ where the notation for angular momentum is standard and $i=1\,(8)$ for color-singlet (color-octet) states.  The short distance cross sections are perturbatively calculable in a power series in $\alpha_s$, while the LDMEs are nonperturbative and must be extracted from data. The LDME scale with definite powers of $v$ so the NRQCD factorization formalism  organizes 
 the calculation of quarkonium production (and decay) into a systematic double expansion in $\alpha_s$ and $v$. 

For $J/\psi$ production, the leading matrix element in the $v$ expansion is $^3S_1^{[1]}$ which scales as 
$v^3$. The next most important are the color-octet  LDMEs: $^3S_1^{[8]}$, $^1S_0^{[8]}$, and $^3P_J^{[8]}$, which all scale as $v^7$.
 $J/\psi$ production has been measured in a wide variety of experiments, including $e^+e^-$, $pp$, $p\bar{p}$, $ep$, $\gamma p$, and $\gamma \gamma$ collisions,
 spanning a wide range of energies. At present, next-to-leadiing order (NLO) QCD calculations are available for the above mentioned color-singlet and color-octet mechanisms for all these initial states.
Global fits to the world's data using these calculations were performed in Refs.~\cite{Butenschoen:2011yh,Butenschoen:2012qr}. The resulting LDMEs are shown in the first line in Table \ref{tb:ldme}. The LDMEs are consistent with the expected $v^4$ suppression of the color-octet mechanisms. The global fits in Refs.~\cite{Butenschoen:2011yh,Butenschoen:2012qr} are reasonably well described by NLO NRQCD, but there are nagging discrepancies that call into question our understanding of quarkonium production.
The most notable discrepancy is the polarization puzzle: if the LDMEs of Refs.~\cite{Butenschoen:2011yh,Butenschoen:2012qr} are used, the produced $J/\psi$
 are predicted to be polarized transverse to their momentum at high  $p_T$, while in fact they are are produced with essentially no polarization. (This is also a problem in $\Upsilon$ production.) Another important discrepancy is the failure of spin symmetry predictions for $\eta_c$ production~\cite{Aaij:2014bga,Butenschoen:2014dra}.
(For possible solutions to the $\eta_c$ problem using different extractions of LDMEs see ~\cite{Han:2014jya,Zhang:2014ybe,Sun:2015pia}.) In light of the failure of NLO QCD to predict the $J/\psi$ polarization, other authors have proposed alternative approaches to fitting the LDMEs. Refs.~\cite{Chao:2012iv,Bodwin:2014gia} have emphasized that NRQCD factorization should be most reliable at the highest values of $p_T$ and have performed fits that focus exclusively on high $p_T$ $J/\psi$ production in colliders. Ref.~\cite{Bodwin:2014gia} also merges NLO calculations with fragmentation contributions in which Altarelli-Parisi evolution is used to resum logs of $p_T/m_{J/\psi}$. 
The LDMEs from the fits of Refs.~\cite{Chao:2012iv,Bodwin:2014gia} are shown in the second and third lines of Table \ref{tb:ldme}, respectively. We will use these three sets of LDMEs in our analysis. There have been other fits to the LDMEs~\cite{Gong:2012ug, Bodwin:2015iua} which include explicit feeddown from $\chi_{cJ}$ states. Since these effects are not included in our calculations, we do not use these LDME extractions in this work. 

\begin{table*}[t!]
\begin{center}
\begin{tabular}{|l|r|r|r|r|r|}
\hline
 & $\langle\mathcal{O}^{J/\psi}(^3S_1^{[1]}) \rangle$  & $\langle \mathcal{O}^{J/\psi}(^3S_1^{[8]}) \rangle $ & $\langle \mathcal{O}^{J/\psi}(^1S_0^{[8]}) \rangle$ & $\langle \mathcal{O}^{J/\psi}(^3P_0^{[8]}) \rangle/m_c^2 $ \\
& $\times$ GeV$^3$ & $\times 10^{-2}$ GeV$^3$ &  $\times 10^{-2} $GeV$^3$ & $\times10^{-2} $GeV$^3$ \\
\hline \hline
B \& K \cite{Butenschoen:2011yh, Butenschoen:2012qr}& $1.32 \pm 0.20$ & $0.224 \pm 0.59$ & $4.97 \pm 0.44$ & $-0.72 \pm 0.88$\\
\hline
Chao, et al. \cite{Chao:2012iv}& $1.16\pm 0.20$ & $0.30 \pm 0.12$ & $8.9\pm 0.98$ & $0.56 \pm 0.21$ \\
\hline
Bodwin et al. \cite{Bodwin:2014gia} & $1.32\pm 0.20$ & $1.1\pm 1.0$ & $9.9 \pm 2.2$ & $0.49 \pm 0.44$ \\ 
\hline
\end{tabular}
\end{center}
\caption{LDMEs for NRQCD production mechanisms used in this paper in units of ${\rm GeV^3}$.  }
\label{tb:ldme}
\end{table*}

Recently, Ref.~\cite{Baumgart:2014upa} proposed studying the distribution of quarkonia produced within jets as an alternative test of NRQCD in hadron colliders.  Cross sections for jets with identified 
hadrons are given in terms of fragmenting jet functions (FJF) that were first introduced in Ref.~\cite{Procura:2009vm} and studied further in 
Refs.~\cite{Liu:2010ng,Procura:2011aq,Jain:2011iu,Jain:2011xz,Jain:2012uq,Bauer:2013bza, Ritzmann:2014mka,Kaufmann:2015hma, Kang:2016ehg,Kang:2016ioz,Dai:2016hzf,dai:2017dpc}. The FJFs are functions of the jet energy, $E$, and the fraction of energy carried by the identified hadron, $z$. FJFs are calculable as a convolution of the more inclusive fragmentation function with a perturbative matching coefficient evaluated at the jet energy scale, $E_J = 2 E \tan(R/2)$. Ref.~\cite{Baumgart:2014upa}
showed that the quarkonium FJF can be calculated in terms of NRQCD fragmentation functions~\cite{Braaten:1994vv,Braaten:1993rw,Braaten:1993mp} and that the $z$ and $E$ dependence of these cross sections 
are sensitive to the underlying production mechanisms because NRQCD fragmentations differ for different production mechanisms. 
For further work see Refs.~\cite{Bain:2016clc, Bain:2016rrv,Chien:2015ctp}.

Ref.~\cite{Bain:2016clc} used the FJF formalism to compute  cross sections for jets with $B$ mesons and $J/\psi$ produced within jets in $e^+e^-$ collisions.
For $B$ mesons the paper studied $e^+e^- \to b\bar{b}$ followed by $b$ quark fragmenting to a jet with a $B$ meson. For $J/\psi$, Ref.~\cite{Bain:2016clc}
studied $e^+e^-\to b \bar{b}g$ followed by gluon fragmentation to a jet with $J/\psi$. In both cases Ref.~\cite{Bain:2016clc} studied the dependence 
of the cross section on $z$, the fraction of the energy carried by the identified hadrons, and the jet's angularity, $\tau_a$,~\cite{Berger:2003iw} a jet substructure variable whose definition can be found in Ref.~\cite{Bain:2016clc}. The analytic expression for these cross sections has the schematic form
\bea
d\sigma[e^+e^-\to jets, h] = H \otimes S \otimes J (\otimes J) \otimes {\cal G}^h
\label{factthm}
\eea
where $h$ is either a $B$ meson or $J/\psi$, ${\cal G}^h$ is the FJF for the hadron $h$, $J$ is a jet function for the other jets in the event (there is one $J$ for $B$ mesons and two for $J/\psi$), $S$ is the soft function, and $H$ is the hard cross section for $e^+ e^- \to b\bar{b}$ or $e^+ e^- \to b\bar{b}g$. Dependence on all kinematic quantities has been suppressed. Evaluation of each of the quantities appearing in Eq.~(\ref{factthm}) shows that they all have logarithms of different scales. The renormalization group equation (RGE) for each of the functions 
in Eq.~(\ref{factthm}) needs to be solved, and the functions need to be evolved to a common scale so that large logarithms are resummed to all orders in perturbation theory. 
Details of this formalism for jet cross sections can be found in  Ref.~\cite{Ellis:2010rwa}. Analytic calculations in Ref.~\cite{Bain:2016clc} were performed to next-to-leading-log-prime (NLL') accuracy.\footnote{NLL' means that in addition to NLL accuracy, $O(\alpha_s)$ contributions to the soft, jet, and FJF are also kept.} The calculations were also performed 
using the PYTHIA monte carlo. For $B$ mesons PYTHIA and NLL' analytical calculations were in good agreement. However, for jets with $J/\psi$ 
good agreement was found only in the $\tau_a$ distributions. The $z$ distributions predicted by PYTHIA at LO were significantly harder than the $z$ distributions predicted by the NLL' calculations. This discrepancy between theoretical calculations and PYTHIA is remarkably similar to the discrepancy between data and PYTHIA recently found by the LHCb collaboration. This motivates us to perform calculations similar to those of Ref.~\cite{Bain:2016clc} to obtain a better description of the LHCb data.

To understand the discrepancy between analytical NLL' calculations of the $z$ distributions and PYTHIA, one must understand how PYTHIA models the production of quarkonium.
In PYTHIA the heavy quark-antiquark pair is produced in the short-distance process in either a color-octet or color-singlet state. If it is in a color-singlet state the heavy quark antiquark pair behaves like a color-singlet particle, emits no gluon radiation, and eventually turns into the quarkonium. If the heavy quark-antiquark pair is produced in a color-octet state, PYTHIA treats the pair as a single colored particle that showers with the splitting function $2 P_{qq}(z)$. As this splitting function is strongly peaked at $z=1$, the color-octet pair retains most of its momentum after the shower. At the end of the shower the color-octet quark antiquark emits a soft gluon in order to become a color-singlet quarkonium. 

The physical picture of quarkonium production implied by the NLL' analytic calculations in Ref.~\cite{Bain:2016clc}  is quite different from PYTHIA. The FJF that controls the $z$ dependence of the cross section is, up to $O(\alpha_s(E_J))$ corrections, equal to the fragmentation function evaluated at the scale $E_J$. The evolution of the fragmentation function from the scale $2 m_c$ up to the 
scale $E_J$ is governed by Altarelli-Parisi evolution equations. Ref.~\cite{Bain:2016clc} showed that this is equivalent to producing a hard gluon in the short-distance  process with virtuality of order $E_J$, allowing that gluon to shower until a gluon with virtuality $\sim 2 m_c$  hadronizes into the $J/\psi$. This can be implemented in PYTHIA by simulating events 
in which the gluon is produced in the hard process, hadronization is turned off, and allowing the parton to shower down to a scale $\sim 2 m_c$. After this a gluon $z$ distribution is obtained and convolved manually with a perturbative NRQCD fragmentation function (calculated at lowest order in $\alpha_s(2 m_c))$. This procedure was referred to Gluon Fragmentation Improved PYTHIA (GFIP) in Ref.~\cite{Bain:2016clc}, and was shown to give good agreement with the NLL' analytic calculation.

\begin{figure}[!t]
\begin{center}
\includegraphics[width=8cm]{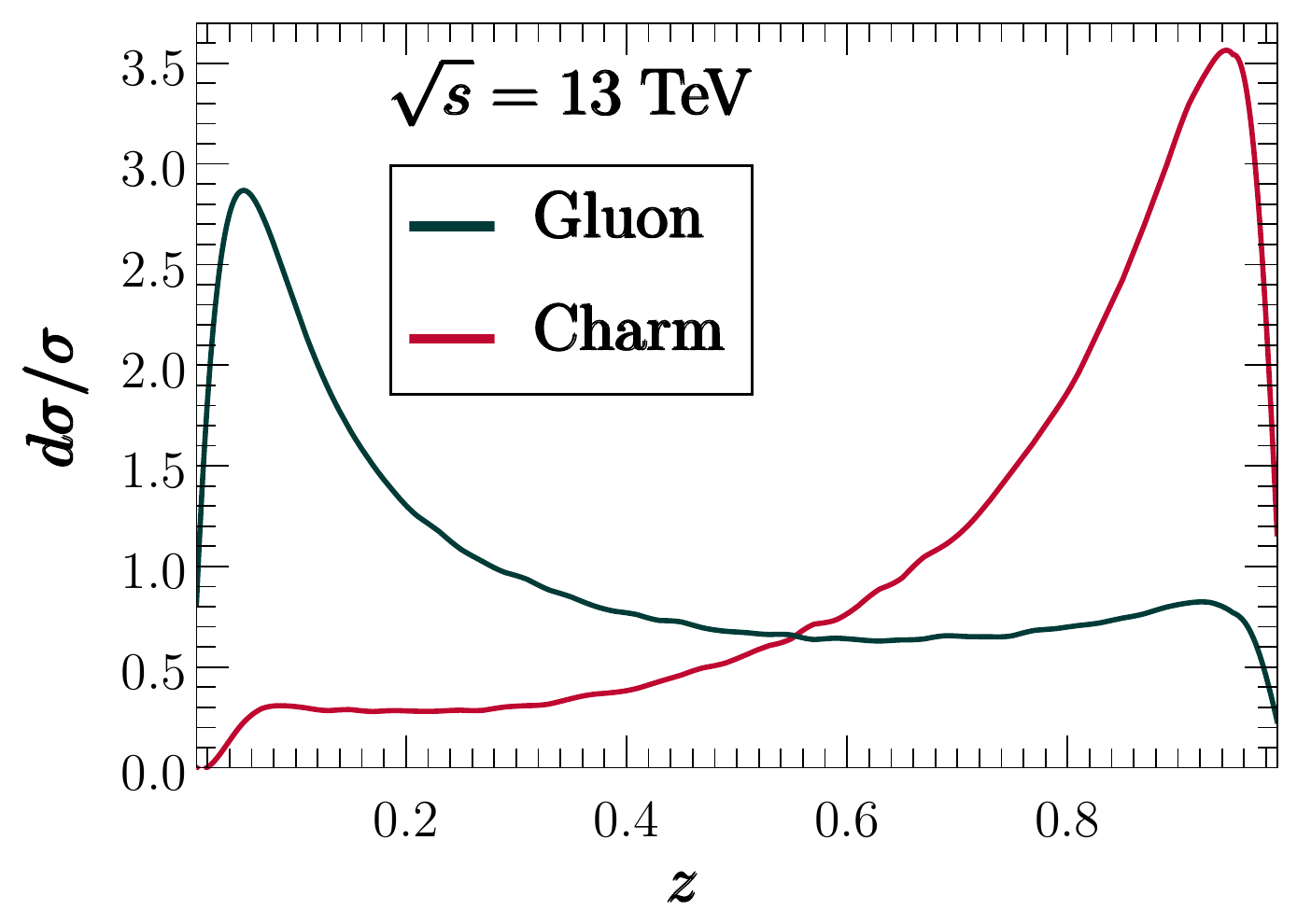}
\end{center}
\vspace{- 0.85 cm}
\caption{
\label{fixedz} 
\baselineskip 3.0ex
PYTHIA predictions for $c$ quark and gluon $z$ distributions (where $z$ is the fraction of the energy of the parton initiating the jet) 
after showering to the scale $2 m_c$. }
\end{figure}

In this letter we perform the corresponding calculation for the LHCb experiment using two different methods. Our first method, which we will refer to as GFIP, is analogous to the GFIP calculation of Ref.~\cite{Bain:2016clc}. We start by generating events corresponding to hard production of $c$ quarks and gluons in $pp$ collisions at $\sqrt{s}=13$ GeV using MadGraph~\cite{Alwall:2014hca}.\footnote{Contributions to $J/\psi$ production from quarks other than $c$ in the hard process are suppressed, either due to soft gluon emission or by $\alpha_s$ evaluated at a large energy scale. We therefore we neglect their contribution.} In the LHCb data, all jets have pseudorapidity $2.5< \eta < 4.0$, $R=0.5$, and the jets are required to have $p_T > 20$ GeV. The hard partons generated by MadGraph satisfy the jet constraints of LHCb. PYTHIA is then used to shower the event down to a scale of $\sim 2 m_c$. 
Jet algorithms are applied to the output of the PYTHIA shower 
and the $c$ quarks and gluons must be within jets of radius  $R=0.5$ satisfying the criteria of the LHCb data described above. The resulting 
$c$ and gluon distributions are shown in Fig.~\ref{fixedz}. Note that the $c$ quark distribution is peaked near $z=1$ while the gluon $z$ distribution is much softer and peaked near  $z=0$.

The $p_T$ and $y$ distributions for the $c$ quarks and gluons are then 
convolved manually with the  NRQCD fragmentation functions evaluated at leading order (LO) in perturbation theory to obtain $p_T$ and $y$ distributions for $J/\psi$. For gluons we include $^3S_1^{[1]}$, $^3S_1^{[8]}$, $^1S_0^{[8]}$ and $^3P_J^{[8]}$ fragmentation functions, because  
the $v^4$ suppression of the color-octet LDMEs is compensated by powers of $\alpha_s$ for $^1S_0^{[8]}$ and $^3P_J^{[8]}$ and $\alpha_s^2$ for $^3S_1^{[8]}$. 
See Ref.~\cite{Baumgart:2014upa} for the explicit expressions for the LO NRQCD fragmentation functions. Color-singlet and color-octet fragmentation functions start at the same order in $\alpha_s$ for charm quarks so we include only color-singlet fragmentation for charm quarks. 
LHCb requires both muons have $2.0 < \eta < 4.5$, $p > 5$ GeV, and $p_T> 0.5$ GeV.  The energy cut clearly suppresses contributions from partons with low $z$ and hence enhances the contribution from $c$ quark initiated jets. We implement the muon cuts by assuming the $J/\psi$ are unpolarized and therefore decays to $\mu^+\mu^-$ isotropically in its rest frame, and the LHCb cuts on the muons are applied to the muons after they are boosted back to the lab frame. From this a normalized distribution in $z(J/\psi)$ is constructed for each production mechanism. Each mechanism is characterized by an initial parton $i$ and quantum numbers $n$, and is multiplied by a weight  
\bea\label{weight}
r(i,n) = \frac{d\hat\sigma(pp\to i+X) \int_0^1dz D^{n}_{i \to J/\psi}(z)}{d\hat \sigma(pp\to c+X) \int_0^1dz D^{^3S_1^{[1]}}_{c \to J/\psi}(z)} \, .
\eea
The weight in Eq.~(\ref{weight})
ensures that the total number of $J/\psi$ coming from each mechanism are in the proper ratio where $D_{i \to J/\psi}^n(z)$ are calculated at the scale
$2m_c$. This is where the fitted LDMEs enter the calculation as $D_{i\to J/\psi}^n(z) \propto \langle {\cal O}^{J/\psi}(n)\rangle$. 
The LHCb data is normalized so that the sum of the heights of the bins adds to 1. Because of 
possible large corrections near $z\to0$ and $z\to1$, we only compare with LHCb data in the range $0.1 < z <0.9$ and normalize our distributions to the sum of the data in these bins.

\begin{figure}[!t]
\begin{center}
\includegraphics[width=8 cm]{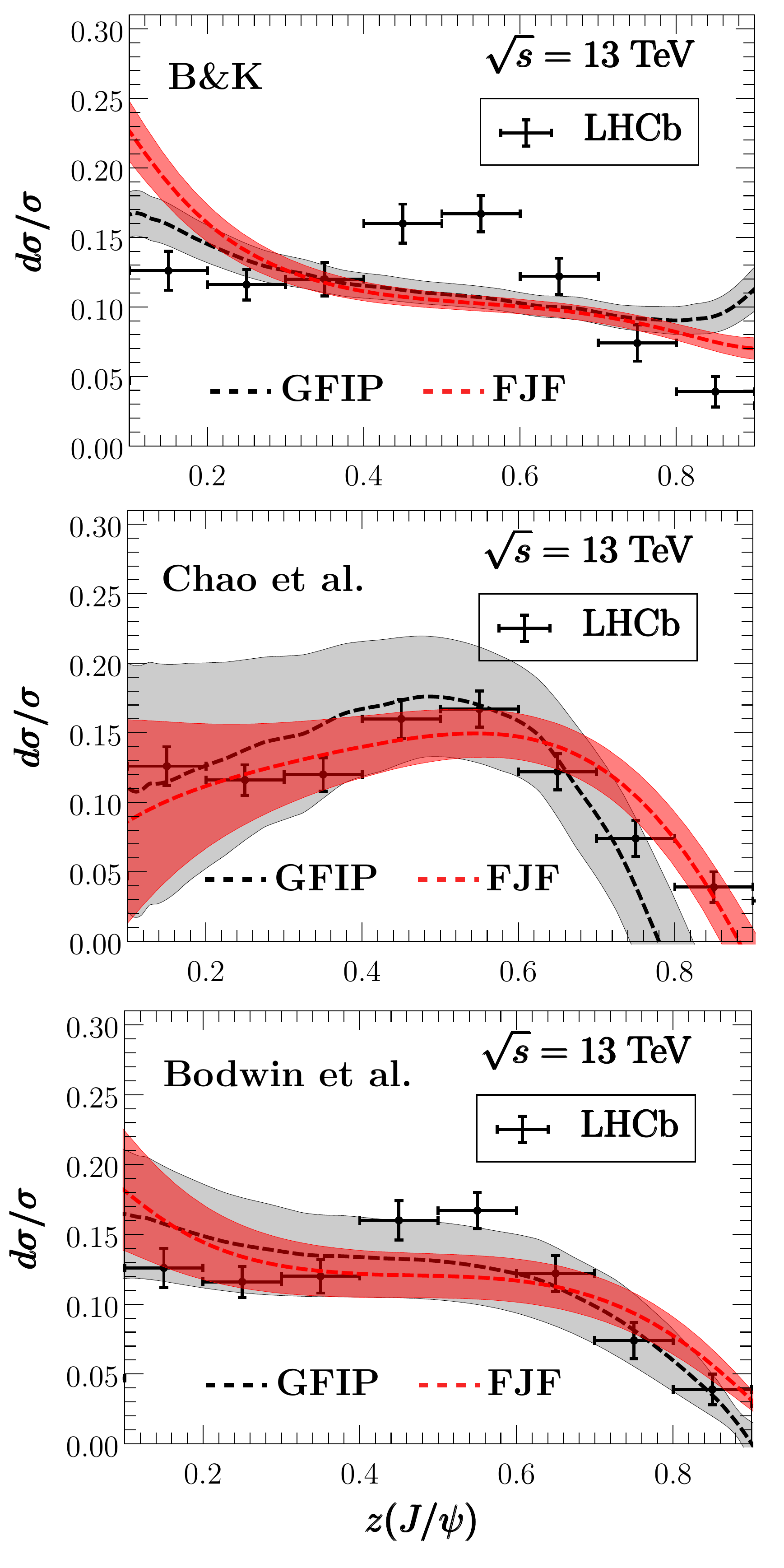}
\end{center}
\vspace{- 1.0 cm}
\caption{
\label{plotsGFIP} 
\baselineskip 3.0ex
Predicted $z(J/\psi)$  distribution using GFIP (gray) and FJF (red) for the three choices of LDME in Table 1 and the LHCb measurements of $z(J/\psi)$. }
\end{figure}

Our second method, which we refer to as the FJF method, employs FJFs  combined with  hard events generated by Madgraph at LO.
In calculating the FJFs,  logartihms of $m_{J/\psi}/E_J$ are resummed using leading order DGLAP equations to evolve the fragmentation functions from the scale $2m_c$ to the jet energy scale, $E_J$.  Madgraph calculates the remaining terms in the factorization theorem to LO in perturbation theory. 
This does not include NLL' resummation for the remaining terms in the factorization thereom, 
however the $z(J/\psi)$ dependence of the cross section is controlled primarily by the FJF.
 The  energy distribution of hard partons is combined with the FJFs for 
 anti-$k_T$ jets~\cite{Waalewijn:2012sv} with $R=0.5$ to produce a $z(J/\psi)$ distribution for each of the five mechanisms. 
 From the GFIP calculations, we know as a function of $z$ the fraction of $J/\psi$ that survive the muon cut and we apply this correction to our 
 analytic calculations. The $z(J/\psi)$ distributions from each mechanism
are weighted by the factors in Eq.~(\ref{weight}) as before.  
The FJF is appropriate for $n$-jet cross sections like Eq.~(1).  Inclusive FJFs \cite{Kang:2016ehg,Kang:2016ioz,Dai:2016hzf,dai:2017dpc} differ
by a contribution from  out-of-jet radiation  that is power suppressed for $R\sim O(1)$~\cite{Ellis:2010rwa}.

Fig.~\ref{plotsGFIP} shows the  predicted $z(J/\psi)$ distributions for the three choices of LDME's in Table \ref{tb:ldme} using the GFIP (gray) and
FJF (red) methods, which are in good agreement. Uncertainties are due to the LDMEs only.  In the case of Ref.~\cite{Bodwin:2014gia}, the errors in
Table \ref{tb:ldme} are supplemented with an error correlation matrix  \cite{Bodwin}. In Ref.~\cite{Chao:2012iv} a fixed relationship between the
$^3S_1^{[8]}$ and $^3P_J^{[8]}$ LDMEs is required to obtain unpolarized $J/\psi$. This constraint is taken into account when computing the uncertainty
due to the LDMEs. These constraints significantly reduce the uncertainty in the predictions relative to naively adding uncertainties in Table
\ref{tb:ldme} in quadrature.  Other sources of uncertainty such as scale variation have not been included. Estimating theory uncertainties reliably in
the absence of a complete factorization theorem is difficult. For example, using the FJF method, the $\mu$ dependence of the FJF should be cancelled by
$\mu$ dependence in hard and soft functions that have not been computed. Note that since the normalization of theoretical curves is fixed to the LHCb
data, any scale variation that affects normalization but not the shapes of the $z(J/\psi)$ distribution will not contribute to the uncertainty.
Especially at low values of $z$, the  underlying event and double parton scattering give additional theoretical uncertainties. 
However, it is not clear how estimate  these uncertainties.

All three choices of LDMEs give better agreement to the LHCb data than default PYTHIA shown in Ref.~\cite{Aaij:2017fak}. This gives support to the picture of quarkonium production  in Ref.~\cite{Bain:2016clc} and this letter. The LDMEs from global fits~\cite{Butenschoen:2011yh, Butenschoen:2012qr} give worse agreement than the fits from Refs.~\cite{Bodwin:2014gia,Chao:2012iv}. The LHCb data is a decreasing function of $z(J/\psi)$ as $z(J/\psi) \to 1$.
This is a property of the $^3S_1^{[1]}$ and $^1S_0^{[8]}$ FJFs, but not the $^3S_1^{[8]}$ and $^3P_J^{[8]}$ FJFs, which actually diverge as $z\to1$. In order to obtain negligible polarization at high $p_T$, the $^3S_1^{[8]}$ and $^3P_J^{[8]}$ LDMEs of Refs.~\cite{Bodwin:2014gia,Chao:2012iv} have relative signs such that they roughly cancel, so the $^1S_0^{[8]}$ dominates production and $J/\psi$ are unpolarized. The same cancellation here allows the $z(J/\psi)$ distribution go to zero as $z(J/\psi)\to1$. Such a cancellation does not occur for the LDMEs from the global fits so the $z(J/\psi)$ distribution starts to turn up at large $z(J/\psi)$.

To summarize, we have analyzed the recent LHCb data on $J/\psi$ production within jets. We used a combination of Madgraph, PYTHIA, and LO NRQCD fragmentation 
functions first introduced in Ref.~\cite{Bain:2016clc} as well as an approach based on Monte Carlo evaluation of the hard process combined with $J/\psi$ FJFs evaluated
at the jet energy scale. Both methods yield $z(J/\psi)$ distributions that agree much better with data than default PYTHIA simulations. The $z(J/\psi)$ distributions are 
very well described by LDMEs from fits to  large $p_T$ data, and less well described by LDMEs from global fits. It would be interesting to perform a combined fit to the LHCb data and the large $p_T$ data used in Refs.~\cite{Bodwin:2014gia,Chao:2012iv} to see if consistent LDMEs with smaller errors can be obtained. Experimental measurement of jets at central rapidity and  the polarization of $J/\psi$ as a function of $z(J/\psi)$ \cite{Kang:2017yde} would also be of interest. Finally it would be especially interesting to find ways of discriminating 
charm and gluon initiated jets \cite{Ilten:2017rbd}, as a sample containing only gluon initiated jets will have greater sensitivity to color-octet LDMEs.

\begin{acknowledgments}
The authors would like to thank P. Ilten and M. Williams for correspondence during the completion of this work.
 RB, YM, and TM are supported in part by the Director, Office of Science, Office
of  Nuclear  Physics,  of  the  U.S.  Department  of  Energy  under  grant  numbers  DE-FG02-05ER41368.  RB is supported by a National Science Foundation Graduate Research Fellowship under Grant No. 3380012. 
AL and LD were supported in part by NSF grant PHY-1519175. 
\end{acknowledgments}
\bibliographystyle{apsrev4-1}

\bibliography{paper}

\end{document}